# DETECTION OF AN ANCIENT PRINCIPLE AND AN ELEGANT SOLUTION TO THE PROTEIN CLASSIFICATION PROBLEM

ASHOK PALANIAPPAN

Madras Institute of Systems Biology®
Chennai 600040
India
Phone: +91 44 2618775

**ABSTRACT**

This work is involved with the development of a well-founded, theoretically justified, and least complicated metric for the classification of proteins with reference to enzymes. As the signature of an enzyme family, a catalytic domain is easily fingerprinted. Given that the classification problem has so far seemed intractable, a classification schema derived from the catalytic domain would be satisfying. Here I show that there exists a natural *ab initio* if nonobvious basis to theorize that the catalytic domain of an enzyme is uniquely informative about its regulation. This annotates its function. Based on this hypothesis, a method that correctly classifies $K^+$-ion channels into their respective subfamilies is described. To put the principle on firmer ground, extra validation was sought and obtained through co-evolutionary analyses. The co-evolutionary analyses reveal a departure from the notion that $K^+$-ion channel proteins are functionally modular. This finding is discussed in light of the prevailing notion of 'domain'. These studies establish that significant co-evolution of the catalytic domain of a gene with its conjoint domain(s) is a specialized, necessary process following fusion and swapping events in evolution. Instances of this discovery are likely to be found pervasive in protein science.

# INTRODUCTION

The quest for principles of biological classification began with the Greek philosopher Aristotle and culminated with Linnæus' binomial system. The problem of functional classification of proteins is recognized as a principal objective of computational biology. Excellent efforts have addressed the problem with a variety of increasingly sophisticated methods (for a review, see [1]). Computational methods for protein classification have the potential to transform the design of biochemical experiments. Function annotation resources have different areas of optimum application owing to the differing strengths and weaknesses of their underlying methodologies. The interested reader is referred to [2] for a discussion of this issue. In summary, current methods are not adequate in the prediction of subfamilies, and a principle of classification rooted in biology is desired.

# A NEW PRINCIPLE OF CLASSIFICATION

I propose as an answer a system of classification founded on the degree of similarity in the domain of catalysis. The reasoning is best illustrated in the philosophical tradition. Enzyme activity is regulated by its biochemical pathway. This implies mediation between enzyme domains and corresponding regulating factors through specific physical interactions. The set of these interactions fixes the control for enzyme activity. Without a robust communication between the regulatory apparatus and the enzyme catalytic domain, regulation would be ineffective. Such mediation surfaces are formed by matching contributions from the catalytic and the regulatory domains. Note that this interaction must be quite specific; so that a catalytic domain is adapted to respond (optimally) to its regulatory factor. Given the variety of possible regulation, individual variations in the catalytic domain necessary to achieve effective interfaces between catalysis and regulation may serve to uniquely identify the protein, i.e., annotate its function. *The catalytic domain enciphers the subfamily of the protein*. The author's discovery of this odd principle presents the immediate application of a wide-ranging solution to the protein classification problem. If two catalytic domains of the same protein family are regulated by similar factors, they must themselves be similar. They must be different if regulated by different factors.

It remains to choose a mathematical technique that is sensitive to the variations in the catalytic domain. A wide range of techniques are at our disposal; for the issues here, we restrict discussion to one such technique, the neighbor-joining method of making evolutionary trees. The neighbor-joining method requires as input an algorithm for calculating the distance matrix of the "objects". For the benefit of a few readers, a discussion of this technique is provided [3].

An object is an Operational Taxonomic Unit (OTU), and a pair of 'neighbors' is a pair of OTUs connected through a single interior node in an unrooted, bifurcating tree. The number of pairs of neighbors in a tree depends on the tree topology. For a tree with N OTUs, the minimum number is always two, whereas the maximum number is N/2 when N is an even number and (N - 1)/2 when N is an odd number. Consider three related OTUs, namely OTU 1, OTU 2, and OTU 3. If we combine, say, OTUs 1 and 2, this combined OTU (l-2) and OTU 3 become a new pair of neighbors. It is possible to define the topology of a tree by successively joining pairs of neighbors and producing new pairs of neighbors. In general, (N – 2) pairs of neighbors can be produced from a bifurcating tree of N OTUs. By finding these pairs of neighbors successively, we can obtain the tree topology.

In general we do not know which pairs of OTUs are true neighbors. Therefore, the sum of branch lengths ($s_{ij}$) is computed for all pairs of OTUs, and the pair that shows the smallest value of $s_{ij}$ is selected as a pair of neighbors. In practice, even this pair may not be a pair of true neighbors; but, for a purely additive tree with no backward and parallel substitutions, this method is known to choose pairs of true neighbors. Thus, if $s_{12}$ is found to be smallest among all $s_{ij}$ values, OTUs 1 and 2 are designated as a pair of neighbors, and these are joined to make a combined OTU (l-2). The distance between this combined OTU and another OTU j, $D_{(1-2)j}$, is given by:

$$D_{(1-2)j} = \frac{1}{2}(D_{1j} + D_{2j}) \forall 3 \leq j \leq N \tag{1}$$

The OTU $j$ with the smallest $D_{(1-2)j}$ is designated as the neighbor of OTU(1-2), giving a new combined OTU(1-2)j. Thus, the number of OTUs is reduced by one, and, for the new distance matrix, the above procedure is again applied to find the next pair of neighbors. This cycle is repeated until the number of OTUs becomes three, which completes the algorithm. At this point, there is only one unrooted tree, which is the desired tree.

**SELECTION OF TEST CASE**

It is possible to set a high benchmark of validation by a limited testing of the hypothesis, provided we take care. The desirable criteria are that the protein family is large, diverse, and membrane-localised. It is easy to identify whether a given sequence could be a membrane protein. This is because membrane proteins are sufficiently hydrophobic to span the lipid bilayer, and the sequence of a membrane protein must have a dominant hydrophobic character. It is easier to predict the secondary structure topology of membrane proteins than that of soluble proteins. This feature owes to the fact that most membrane proteins spontaneously form backbone hydrogen bonds upon insertion into the membrane and adopt α-helical transmembrane structures. (The exceptions are membrane proteins composed of β-barrel; e.g., bacterial porin channels.) It would seem altogether surprising that the 3-D structure of membrane proteins is much harder to obtain than that of soluble proteins. Given this difficulty, a system of classification that has been tested with globular proteins may not be successful with the class of membrane proteins. The truth of the converse is open to discussion. We also note that the development of classification methods for specifically annotating membrane proteins has attracted less effort. The characterization of ion channel membrane proteins requires the additional dimension of electrophysiological experiments. The family of potassium ion channel proteins fits the bill perfectly.

The classification problem of $K^+$-ion channels is both complex and important: complex because the physiological diversity of $K^+$-channels precludes complete functional classification with hydropathy analysis; important because $K^+$ channels are the prototypical members of the voltage-gated ion channel superfamily. Their evolution into a huge multigene family was a necessary step in the emergence of complex multicellular organisms. The biophysical classification of $K^+$ channels is based on their electrophysiological and biochemical characteristics. The catalytic domain of $K^+$ channels is the permeation pathway. As expected of catalytic domains, the structure of the permeation pathway is conserved in all life, including ourselves [4-6]. There is evidence that some properties individual to the subfamily are localised on the permeation pathway (see discussion in [7]). These observations consistently interpreted strongly support the candidature of the permeation pathway for underpinning the classification of $K^+$-channels.

**RESULT**

The neighbor-joining tree of the permeation pathway of $K^+$-channels had been derived in the context of an earlier study [2]. In summary, the permeation pathways from eighteen $K^+$-channel subfamilies in the human genome could be sorted into subfamily-ordered monophyletic clusters. Presence of nodes on the tree that radiate subfamilies provides direct evidence that the neighbor-joining tree of permeation pathways accurately classifies $K^+$-channels into their subfamilies.

The success of the above procedure raises a deeper issue: is it a manifestation of co-evolution between the permeation and regulatory domains? I examine this question below and provide complete support for the co-evolutionary argument.

**CO-EVOLUTIONARY ANALYSES**

The measurement of co-evolution between the catalytic domain and the regulatory domain of $K^+$ channels involves:
1. the construction of a suitable dataset;
2. a suitable measure of co-evolution.

To test the co-evolutionary hypothesis, the class of voltage-gated $K^+$ channels (Kv channels) across all phyla was considered, since it is the most extensive and most extensively studied class of K$^+$ channels. The regulatory module is the voltage sensor, S4 helix, which has the pattern R-x-x-R-x-x-R-x-x-R-x-x-[RK] (see fig. 1).

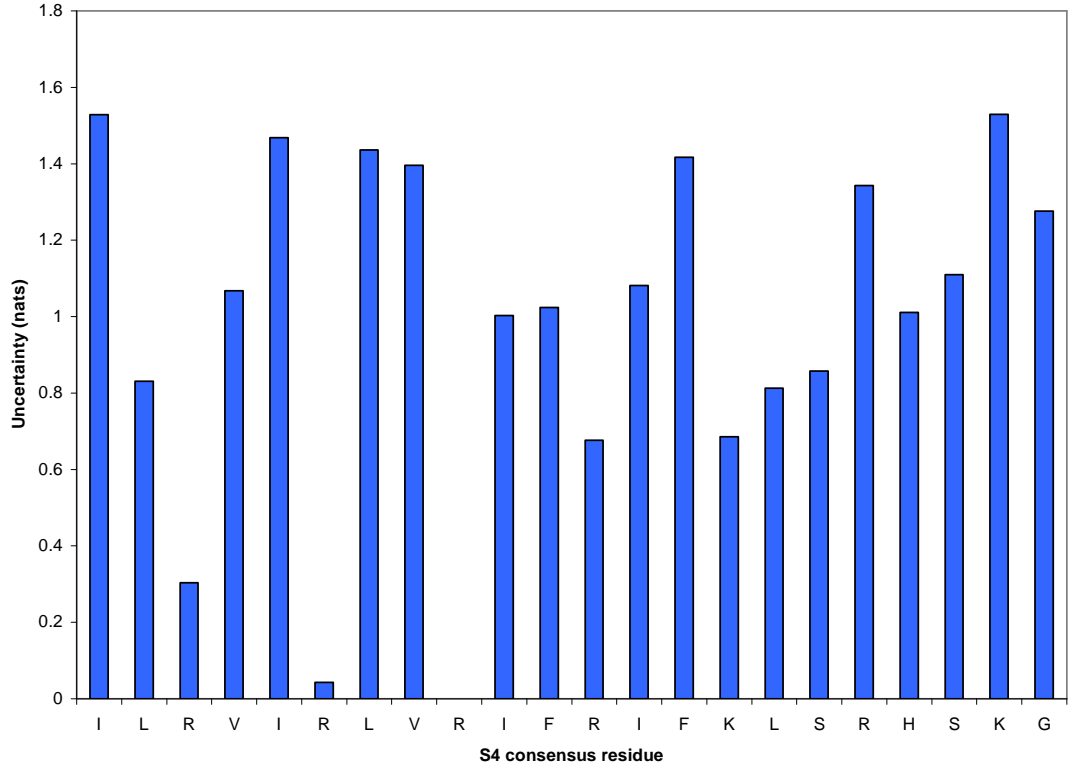

Figure 1. A consensus sequence entropy plot of a 22-residue region representing S4 voltage sensing helix. The plot is constructed of the voltage sensors extracted from a multiple alignment of 147 voltage-gated K$^+$ channels in all life, including eubacteria, archaea, protista, plants and animals. The X-axis represents the consensus residue at that position in the S4 helix. The height of the bars represents the entropy (in nats) along the S4 helix. Every third-spaced residue is positively charged, mostly arginine. There is very small entropy for the second Arg from the start of the S4 helix, and zero entropy in the position of the third Arg. Also observe the population of consensus large hydrophobic residues that filled all the positions available in order to achieve membrane insertion of the S4 helix in spite of the energetic cost caused by ionizable residues [8].

The Kv (i.e., voltage-gated K$^+$) channel dataset was compacted to eliminate redundant sequences in the analysis. I used a 90% sequence identity cutoff, and two sequences that were more than 90% identical were replaced by one of the sequences. Finally I obtained a non-redundant set of 147 Kv channels belonging to one of the following subfamilies: KCNA, KCNB, KCNC, KCND, KCNF, KCNG, KCNS and the Kv sensory channel subfamily. (For a description of these subfamilies, see [2].) The total length of S4 was taken as 22 residues for this analysis. This constituted the gating, regulatory domain in the context of our analysis. For the permeation catalytic domain, we extracted the region following S4 and extending up to the C-terminus of the S6 helix.

Co-evolutionary analysis was originally formulated to identify binding partners between two families. Many proteins must evolve in concert to maintain the energetically and structurally relevant features of a binding interface that they share. The correlated divergent evolution between two families of structurally homologous proteins is defined as co-evolution between the families. For e.g., the divergent evolution of proteins in cellular signaling pathways requires ligands and their receptors to co-evolve, creating new pathways when a new receptor is activated by a new ligand. Variations in the sequence could influence their binding specificity. By relating the sequence similarity of a set of proteins to their binding partner preferences, the binding specificity of an uncharacterized protein can be inferred by its sequence similarity to other characterized proteins within the same family [9].

The measurement of co-evolution is detailed below for the case of two interacting protein domains, namely catalysis and regulation. Evolutionary distance matrices are generated from the multiple alignments of the families of co-evolving protein sequences. In order to quantify the co-evolution of interaction partners, a linear regression analysis measuring the correlation between pairwise evolutionary distances among all catalytic domains and the corresponding pairwise evolutionary distances of their regulatory units is used. X is defined as a two-dimensional matrix of evolutionary distances in the catalytic domain. (X was constructed as a NxN matrix, where N is equal to the number of sequences). For the corresponding regulatory domains, a similar distance matrix, Y was constructed. In particular, $X_{ij}$ signifies the pairwise distance between sequence $m_i$ and sequence $m_j$, and $Y_{ij}$ the pairwise distance between sequence $n_i$ and sequence $n_j$ (where $n_i$ regulates $m_i$ and $n_j$ regulates $m_j$ whether or not they are part of the same polypeptide). The correlation coefficient r between the pairwise distances in matrix X and their corresponding distances in matrix Y is given by:

$$r = \frac{\sum_i \sum_j (X_{ij} - \langle X \rangle)(Y_{ij} - \langle Y \rangle)}{\sqrt{\sum_i \sum_j (X_{ij} - \langle X \rangle)^2 \sum_i \sum_j (Y_{ij} - \langle Y \rangle)^2}} \qquad (2)$$

where $\langle X \rangle$ is the mean of all $X_{ij}$ values and $\langle Y \rangle$ is the mean of all $Y_{ij}$ values. Note that $-1 \leq r \leq 1$. Positive values of r, i.e. a positive coevolution, indicate that an evolutionarily close pair of catalytic domains is regulated by a similarly evolutionarily close pair of factors. That is, the similarity in regulation is reflected in the similarity of the catalytic domains. However, r-values of around zero mean little correlation, whereas negative values of r imply anti-correlation, i.e., similar factors appear to regulate divergent catalytic domains (which is not meaningful evolutionarily speaking). The above method was adapted for measuring the co-evolution of intrasubunit domains of permeation and voltage sensing.

The object of our study is the 'coupling' mechanism that transmits the sensing of voltage from the voltage sensor to the permeation domain, i.e., a quantification of significant correlated changes between the channel regulatory site and the channel activity site. As regulation increases in complexity, the domain of catalysis evolves to keep pace with its partner. Otherwise interaction and hence regulatory function is lost. To effect these complementary changes, a nearly concurrent period of evolution in the catalytic domain takes place as the regulatory domain changes in sequence, structure or function.

Using the dataset constructed above, I constructed a multiple alignment of the permeation catalytic domain and the gating domain ([10]). The regulatory, gating domain (S4) aligned ungapped by virtue of identification. The evolutionary distance matrices of the permeation pathways and the voltage sensors were computed from their respective alignments. The distance matrices were based on percent sequence divergence with correction for multiple substitutions, and computed using the PROTDIST program of the PHYLIP package [11].

Following the above discussion, X was defined as the 147x147 matrix of evolutionary distances of the permeation pathways, and Y is the corresponding distance matrix of the S4 domains. $X_{ij}$ is the pairwise distance between permeation pathway $m_i$ and permeation pathway $m_j$. $Y_{ij}$ signifies the pairwise distance between S4 sensor $n_i$ and S4 sensor $n_j$ (where $n_i$ is *covalently linked* to $m_i$ and $n_j$ is covalently linked to to $m_j$). The correlation coefficient was then calculated for all the pairwise distances in matrix X and their corresponding distances in matrix Y using (2). I obtained r = 0.5065 (see fig. 2). A positive value of correlation is evidence for co-evolution between the permeation

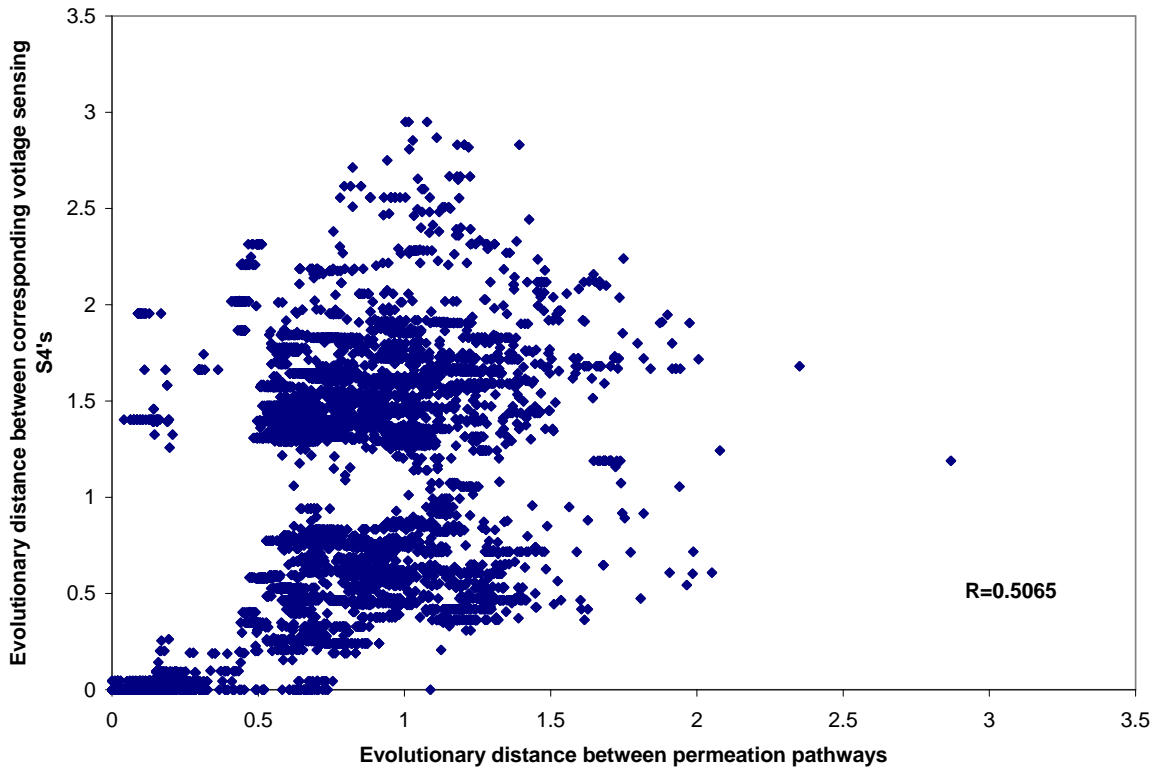

Figure 2. A scatterplot of the permeation pathway evolutionary distances and the voltage sensor evolutionary distances. Each datapoint represents evolutionary distances in the same pair of sequences. The plot is constructed from 18769 datapoints, corresponding to the number of pairs Figure 9 (contd.) reconstructible from a trimmed set of 137 voltage-gated $K^+$ channel sequences. The correlation coefficient of the plot is approximately 0.5 (given on the right side of the figure), implying that there is significant co-evolution between a permeation pathway and its corresponding voltage sensor. Evolutionary distances were estimated using the PROTDIST program of PHYLIP.

pathway and its corresponding S4 for the class of voltage-gated $K^+$ channels.

To draw conclusive inferences, the significance of the correlation must be estimated [9]. The significance of the computed r was assessed by an estimation of the probability of obtaining the observed value of r by chance (p-value). The p-value was obtained by randomly shuffling the pairwise distances between permeation pathways and voltage sensors. Thus the assignments of correspondence (voltage sensor $n_i$ is covalently linked to permeation pathway $m_i$, and voltage sensor $n_j$ is covalently linked to permeation pathway $m_j$) were replaced by random assignments, and the correlation coefficient was computed as above. This process was done 1000 times. (Randomization was implemented using functions available in C.) From the resulting 1000 values of $r_{rand}$, a z-score for the actual observed value r was calculated as:

$$z = \frac{r - \langle r_{rand} \rangle}{\sigma_{rand}} \qquad (3)$$

where $\sigma_{rand}$ is the standard deviation of $r_{rand}$, and $\langle r_{rand} \rangle$ is the mean (effectively zero for truly random data). We obtained a $\langle r_{rand} \rangle = 0.000000$, and $\sigma_{rand} = 0.000005$. This gave a z-score $\approx 1*10^5$ for the correlation coefficient we had calculated for the co-evolution of the permeation pathway with the S4 sensor. The p-value is then computed from:

$$p = erfc(|z|)/\sqrt{2} \qquad (4)$$

where erfc is the complement error function ( $erfc(x) = \frac{2}{\sqrt{\pi}} \int_{x}^{\infty} e^{-u^2} du$ ). I obtained $p << 10^{-6}$.

Thus a p-value analysis of the significance of the correlation coefficient indicated that there was negligible probability of obtaining that high a correlation by chance. This affirmed that there is significant co-evolution between the permeation pathway and the voltage sensor. This p-value was dramatically smaller than those reported for other co-evolutionary analyses ([9]), indicating a more positive co-evolution in our case, also due to the size of our dataset.

Thus we see that the co-evolution between the permeation domain and the regulation domain is robust that it could be detected in a single class of regulation, namely the voltage-gated class. The technique based on permeation pathway would discriminate even better among different modes of regulation, since the adaptation in this context would need to differ *qualitatively*. It follows that the permeation domain optimises interaction with its regulatory factor. This process occurs necessarily independently for each gene, conferring the specificity of co-evolution of the permeation domain. The universality and specificity of the process are the pre-requisites enabling the success of the system of protein classification discussed here.

*Comment on functional modularity:*

The notion of functional modularity is quite widespread in the literature on protein domains. This study prescribes limits to such modularity of function. In the construction of potassium channel chimeras, where the permeation pathway of one $K^+$ channel is spliced with the regulatory domain of another, some channel property would be affected. The change in function is a function of the specific subtypes in the mosaic. This does not contradict the modular history of the gene. We are interested in the co-evolutionary process that took place after catalytic domains acquired new regulatory elements.

## CONCLUDING DISCUSSION OF USEFULNESS OF PRINCIPLE IN PROTEIN CLASSIFICATION

**Function annotation of a new $K^+$ channel.** The extension of the above protocol for the annotation of new $K^+$ channels is straightforward. The selectivity filter is the mark of a $K^+$-channel (see appendix B in [12]). Let the permeation domain of a channel of unknown function be denoted by $j$. Each subfamily already characterized is represented by one collective OTU consisting of permeation domains of the members of the subfamily in topological order. For instance, there are 18 OTUs corresponding to the subfamilies of the human genome. The distances between $j$ and each collective OTU is calculated using (1). The OTU that corresponds to the minimum distance represents the function of the new $K^+$-channel. In the case where no OTU produces a clear minimum, the possibility is considered that the given channel may be an outlier. We may have recorded the first occurrence of a new subfamily. Later this subfamily is included in the repertoire of the protein family as a pre-defined subfamily OTU. In this way, the system is robust to the inclusion of new subfamilies.

**Extension to any protein family.** The generalization of the above language for the classification of any protein family is again straightforward. I have described a validated theory for *in silico* inference of protein function and propounded an organizing principle for the rapid and reliable classification of any given protein family. Sequence annotation is prone to errors, systematic or otherwise, and dependent on sequence sampling (see [13-15]). Pilot studies affirmed the extendability of the method (unpublished). Applicability of function annotation by catalytic domain would be a complementary method to existing technologies. By virtue of its elegance, the principle would greatly accelerate the rate-limiting process of annotation. Its contribution to the automatic analysis of proteomes and streamlining of genome sequencing projects will be considerable.